\documentclass[journal,article,submit,moreauthors,10pt,a4paper]{mdpi} 
\firstpage{1} 
\articlenumber{x}
\doinum{10.3390/------}
\pubvolume{xx}
\pubyear{2016}
\copyrightyear{2016}
\externaleditor{Academic Editor: name}
\history{Received: date; Accepted: date; Published: date}

\newcommand{\enzo}{\it {\small ENZO}}

\Title{ON THE NON-THERMAL ENERGY CONTENT OF COSMIC STRUCTURES}

\Author{Franco Vazza $^{1}$, Denis Wittor $^{1}, $Marcus Br\"{u}ggen$^{1}$, Claudio Gheller $^{2}$}
\address{%
$^{1}$ \quad Hamburger Sternwarte, Gojenbergsweg 112, 21029 Hamburg, Germany; \\ 
$^{2}$ \quad ETHZ-CSCS, Via Trevano 131, CH-6900 Lugano, Switzerland}

\corres{Correspondence: franco.vazza@hs.uni-hamburg.de}

\abstract{1) Background: the budget of non-thermal energy in galaxy clusters is not well constrained, owing to the observational and
theoretical difficulties in studying these diluted plasmas on large scales. 2) Method: we use recent cosmological
simulations with complex physics in order to connect the emergence of non-thermal energy to the underlying evolution of gas and dark matter. 
3) Results: the impact of non-thermal energy (e.g. cosmic rays, magnetic fields and turbulent motions) is found to 
increase in the outer region of galaxy clusters. Within numerical and theoretical uncertainties, turbulent motions dominate the 
budget of non-thermal energy in most of the cosmic volume.  4) Conclusion: assessing the distribution non-thermal energy in galaxy clusters is crucial to perform
high-precision cosmology in the future. Constraining the level of non-thermal energy in cluster outskirts will improve our understanding of the acceleration of 
relativistic particles  and of the origin of extragalactic magnetic fields.}


\conferencetitle{EWASS2016}

\begin{document}



\section{Introduction}

Galaxy clusters are the largest reservoir in the Universe of, both, thermal and non-thermal (NT) energy. They sit atop of the matter hierarchy of the Universe and are still forming nowadays, through the continuous accretion of matter \citep[e.g.][]{2010MNRAS.406.1759M,2012ARA&A..50..353K}.  
Each accretion episode converts a fraction of the infall kinetic energy into gas thermal energy \citep[e.g.][]{ry03,br11}, and is accompanied by the injection of 
turbulent motions across a wide range of scales \citep[e.g.][]{su06,va11turbo,miniati14,bj14,2015ASSL..407..599B}. \\

Mergers can inject an amount of NT energy of the order of the infall kinetic energy, and thus they
can give us  a view onto out-of-equilibrium plasma conditions, where relativistic particles can be 
accelerated \citep[][]{by08,do08} and give rise to observed diffuse radio emission \citep[e.g.][]{fe08,fe12}.
Moreover, a the  NT  pressure support may yield signifiant deviations from  the mass estimates based on the hypothesis of virial theorem and hydrostatic equilibrium, which are inferred from X-ray analysis\citep[e.g.][]{2004MNRAS.351..237R,lau09,2010ApJ...711.1033Z,2013A&A...551A..22}. 
Galaxy clusters are predicted to behave as closed boxes that 
retain all dark and gaseous matter accreted onto their deep gravitational wells. Therefore, observed departures  of their baryon fraction from the cosmic average can inform us
about the relevance of non-gravitational processes, from NT processes to the integrated effect
of galaxy feedback within them \citep[e.g.][]{2006MNRAS.365.1021E,2012ApJ...758...74B, 2013A&A...551A..23}.
 
Also the mass growth rate of clusters, their abundance at a given cosmic epoch and the slope of observed scaling relations between X-ray luminosity, the average temperature and total mass can constrain the cosmological parameters complementary to the analysis of the CMB \citep[e.g.][]{2005RvMP...77..207V,2009ApJ...692.1060V}. 
Therefore, galaxy clusters can be used as ``high-precision'' cosmological tools provided that we achieve a satisfactory understanding of the complex interplay between thermal and NT effects.  This will be important both for a calibration of cluster scaling relations within the standard $\Lambda$CDM model (which  future missions like EUCLID {\footnote{www.euclid-ec.org}}, will probe via gravitational lensing), as well as for the cosmological use of the information encoded in the intergalactic medium, that locks-in $\sim 90$ percent of the baryon in the Universe and will be targeted by future high-fidelity and high-resolution spectrographs, such as those designed for the ELT {\footnote{www.hires-eelt.org}} . \\

Since the NT components should have a broader energy profile compared to the  thermal gas distribution, and are also subject to longer dynamical timescales, they offer the chance to probe the conditions of rarefied cosmic plasma prior to the formation of large-scale structures.

For example, cosmic rays (CR) protons  injected by shocks are long-lived in clusters as they are subject to negligible energy losses and stay confined inside clusters for cosmological timescales  \citep[][]{bbp97,volk99}. Upper limits on the presence of CRs in clusters at low redshift \citep[][]{re03,gb07,fermi14} have the potential to probe the acceleration efficiency of cosmic shocks across the entire lifetime of clusters, and also to constrain the acceleration efficiency of shocks in environments too faint to be directly observable \citep[e.g.][]{va15relics,scienzo16}.

Also magnetic fields in the ICM are subject to very long dissipation timescales. While the dynamical activity in the innermost cluster regions is probably strong enough to boost a small-scale dynamo \citep[e.g.][]{do99,2004ApJ...612..276S,ry08,cho14}, in cluster outskirts and in filaments the dynamo amplification should be much less efficient. Therefore, the dynamical memory of the seed magnetic fields should be still present nowadays in cluster outskirts \citep[][]{va14mhd} and might be detectable by incoming radio telescopes \citep[][]{va15radio}.

In this contribution,  we use state-of-the art cosmological simulations of large-scale structures to assess the spatial distribution of NT energy in the local Universe. Our findings are compared to the scarce observational constraints at the scale of galaxy clusters (Sec. 2.1) , as well as on the scale of large cosmic volumes (Sec 2.2).
Our discussion and conclusions are given in Sec. 3.

\begin{figure}
\centering
\includegraphics[width=0.8\textwidth]{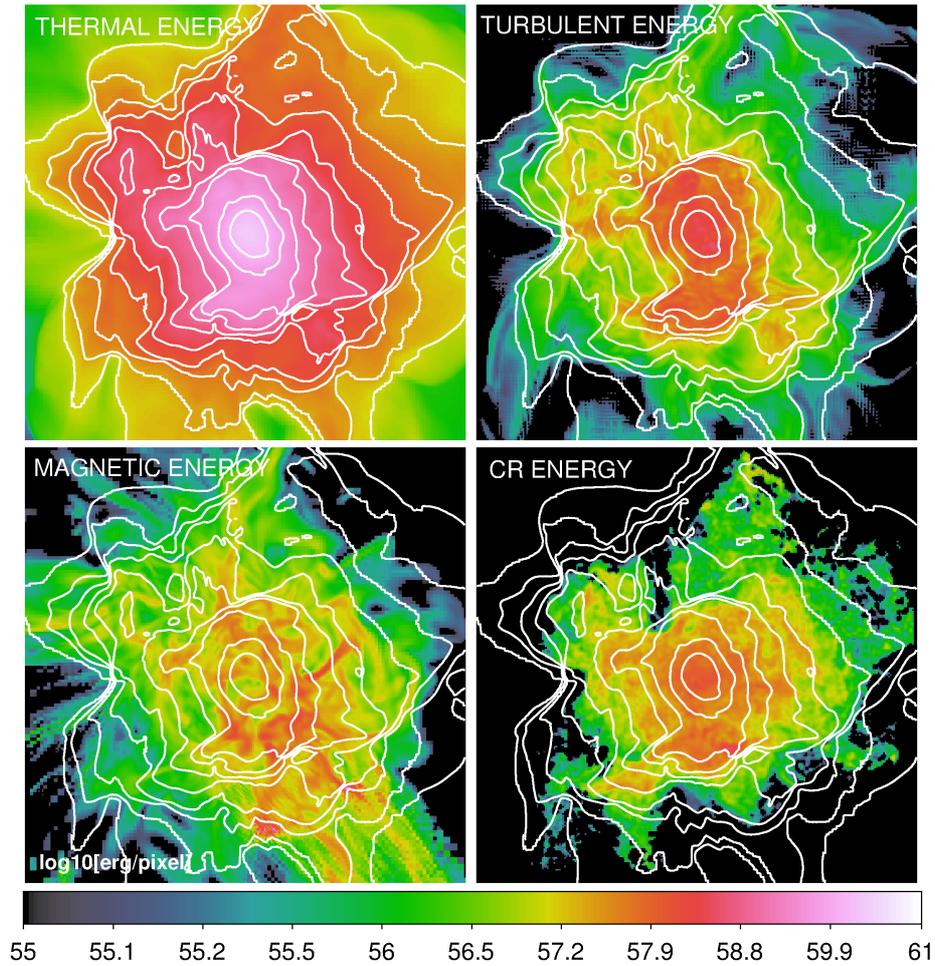}
\caption{Projected energy maps for our simulated clusters at $z=0$, including the 
thermal energy (top left) the turbulent energy (top right), the magnetic energy (lower left) and the CR-proton energy (lower right). The size of each image is $8.192 ~\rm Mpc$. For clarity the white contours of the projected thermal energy are reported in each  panel.}
\label{fig:maps_cluster}
\end{figure}

\section{Results}


\subsection{The distribution of NT energy inside galaxy clusters.}

We first focus on the high-resolution view of one massive ($\approx 1.1 \cdot 10^{15} M_{\odot}$) galaxy cluster. We chose this object because its mass its similar to the one of the Coma cluster, for which many observational constraints are available.  

In detail, we used the grid code {\enzo} \citep[][]{enzo13} and employed 5 levels of adaptive mesh refinement (AMR) to selectively
increase the dynamical resolution in most of the cluster volume, i.e. $\sim 80\%$ of the viral volume is refined up to the maximum resolution of  $\Delta x=31.7 ~\rm kpc$ for $z \leq 1$. 
Our AMR scheme is aggressive in order to allow the largest possible dynamical range in the gas flows, which is crucial to allow the growth of a small-scale dynamo \citep[e.g.][]{2016ApJ...817..127B}:  we refined all cells which are $\geq 10\%$ denser then their surrounding, as well as whenever the 1-D velocity jump with their neighbours is $\geq 1.5$ \citep[e.g.][]{va10kp}. 
More details on this cluster simulation are discussed in \citet{wi16} and Wittor et al. (this volume). 

All our runs followed magneto-hydrodynamics (MHD) using the Dedner Cleaning method of {\enzo} \citep[e.g.][]{wang10,enzo13}.  
We simulated two extreme scenarios for the seeding of magnetic fields: 
a) in the {\it primordial} scenario (used in the non-radiative setup) we started the simulation from a uniform magnetic field seeds of $B_0=10^{-9} \rm G$ (comoving) at $z=30$, which is at the level of upper limits allowed from the analysis of the Cosmic Microwave Background (i.e. $B_0 \leq 0.1-1 \rm ~nG$, e.g. \citealt[][]{sub15,2016A&A...594A..19P}). 
b) In the {\it astrophysical} scenario (combined with the radiative setup, including gas equilibrium cooling and thermal/magnetic feedback from active galactic nuclei) we 
injected magnetic dipoles storing $1 \%$ of the AGN feedback energy, starting from $z=4$.  The thermal feedback is released in bursts of $E_{\rm agn}=10^{60} \rm erg$ thermal energy for events (corresponding to a power of $W_{\rm agn} \approx 5 \cdot 10^{44} \rm erg/s$) , with energy deposited in a couple of over-pressurised outflows at random opposite directions from the halo centre (along one random direction along the coordinate axes). In previous works we showed that this 
simplified approach to include the large-scale effects of AGN in clusters can reasonably well reproduce the innermost thermodynamical profiles of clusters and cluster scaling relations \citep[][]{va13feedback,scienzo16}.  Where the magnetisation from AGN is not present, we initialise a primordial magnetic field to $B_0=10^{-20} \rm G$ (comoving) at $z=30$.


The dynamics of CRs is simulated in post-processing using  tracer particles 
(\citealt{wi16}, see also Wittor et al. this volume). This is a reasonable approximation, under the following assumptions: a) the CRs are frozen into the gas because their spatial diffusion relative to the thermal gas is negligible on our  resolution scale ($\Delta x$); b) they represent a passive fluid with limited influence on the gas dynamics, which is justified by the low, $\sim 1 \%$ energy budget, allowed by $\gamma$-ray upper limits \citep[][]{fermi14} and also based on our previous work using a self-consistent 2-fluid model for CRs \citep[][]{scienzo,scienzo16}. 

In post-processing, we used 240 high-resolution {\enzo}'s snapshots from $z=1$ to $z=0$ to track the propagation of  $\approx 1.33 \cdot 10^7$ tracers using the OPEN-MP code {\it Crater} \citep{wi16}.
Each tracer samples $\approx 10^8 \rm M_{\odot}$ of gas mass, and is monitored in order to model the injection and re-acceleration of CRs, based on an on-the-fly shock finder. CRs are assumed to be injected only by quasi-parallel shocks (angle $\leq 50^{\circ}$ between the shock normal and the upstream magnetic field), based on recent results from particle-in-cell simulations \citep[][]{2014ApJ...783...91C}. For these shocks, the acceleration efficiency as function of Mach number follows from the best acceleration efficiency model found in \citet[][]{scienzo16}, which 
can fulfil the constraints posed by the FERMI satellite \citep[][]{fermi14}. This model is a rescaled version of the model suggested by \citep[][]{kr13}, where the acceleration efficiency is rescaled down $\sim 1/5$, based on recent results by particle-in-cell simulations of quasi-parallel shocks \citep[][]{2014ApJ...783...91C}.

The panels in Figure \ref{fig:maps_cluster} show the projected energy for a $8.192 \times 8.192 ~\rm Mpc^2$ area centred on the cluster centre at $z=0$ (for a depth of $3.17 ~\rm Mpc$ along the line of sight) for a primordial seeding model of magnetic fields. 

The thermal energy $E_{\rm thermal}=3/2 k_{\rm B} \Delta x^3 \rho T/(\mu m_{\rm p})$, where $\rho$ is the gas density within each cell, $T$ is the temperature and $\mu=0.6$ is the mean molecular mass ($k_{\rm B}$ and $m_{\rm p}$ are the Boltzmann constant and the proton mass, respectively). 
The turbulent energy is  $E_{\rm turb}=1/2 \rho \delta v^2 \Delta x^3$, where $\delta v$ is the residual velocity after filtering out velocity structures on scales $L \geq 300 ~\rm kpc$. This scale is chosen for consistency with our previous works \citep[e.g.][]{va11turbo} and is meant to filter out the typical scale of subgroups accreting into clusters of this size. However, the choice of this scale is not unique, as during major merger events the outer scale of turbulent motions can increase, and more complex filtering techniques should be used \citep[][]{va12filter}. In the following, we will also discuss the case in which no large-scale motions are filtered out from the turbulent kinetic energy budget, in order to bracket the real level of turbulence in the ICM. 
The magnetic energy is computed as $E_{\rm mag}=B^2 \Delta x^3/(8 \pi)$, where $B$ is the magnetic field strength in each cell.
Finally, the CR-proton energy in the cell is computed as $E_{\rm cr}= \sum_i e_{\rm cr,i}$, where $e_{\rm cr,i}$ is the CR-energy measured by each tracer particle ending up in the cell; $e_{\rm cr,i}$ is the result of the injection and the reacceleration of CRs by shocks, and of the adiabatic compression/expansion experienced by each tracer,  neglecting energy
losses via Coulomb/hadronic collisions \citep[see][for details]{wi16}.

Our maps show that the thermal gas energy clearly dominates over all other NT energy forms (Fig \ref{fig:maps_cluster}) across the entire cluster volume.  All energy forms roughly follows the gas density profile of the cluster. However, the perturbed dynamical state of the cluster (which underwent a major merger at $z \sim 0.3$) produces a quite patchy and asymmetric distribution of NT energies. In particular, all NT energies are significantly increased in the direction traced by large-scale filaments (approximately from the upper left to the bottom right of each panel), which also is the direction of the last major merger experienced by the cluster. The increase of turbulent and CR-energy density is confined within the merger shocks launched by this merger (marked by the accumulation of thermal energy contour levels in the first panel), while the magnetic energy density has a more extended distribution going out towards the filaments. 

\begin{figure}
\centering
\includegraphics[width=0.8\textwidth]{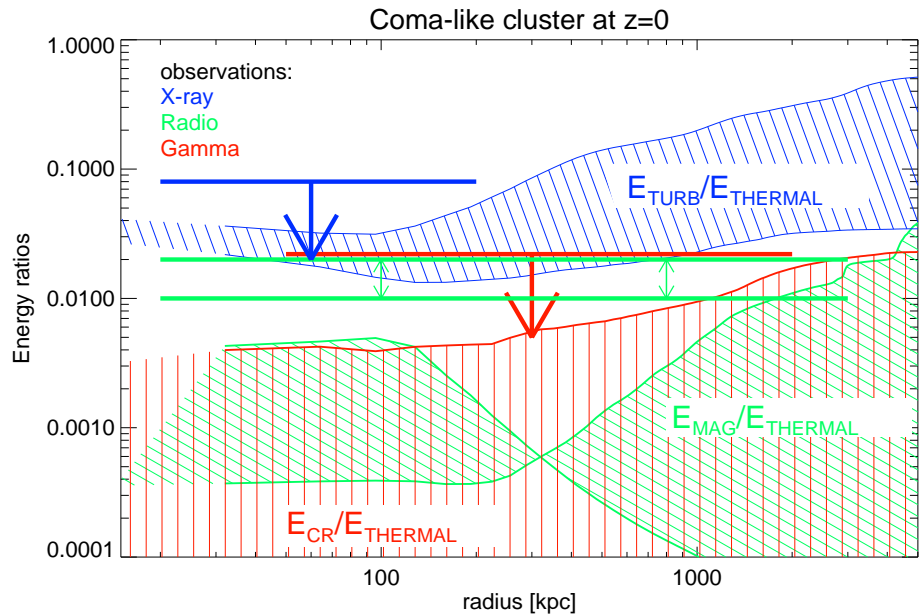}
\caption{Radial profile of the ratios between NT energies and the thermal energy of the ICM for a simulated galaxy clusters at $z=0$.} 
\label{fig:NT_cluster}
\end{figure}

To better quantify the relative contribution of the different kinds of energies, we show in Figure \ref{fig:NT_cluster} the radial profiles of the ratios between each NT energy component and the thermal energy of the ICM. The ratios are computed between {\it enclosed} energies inside the radius (e.g. $\sum_{0\leq i \leq R_i} E_{\rm thermal}(R_i)$, where $E_{\rm thermal}(R_i)$ is the total thermal energy of all cells in each radial shell
with radius $R_i$ and thickness $\Delta R=1$ cell, etc).

The hatching for the profiles brackets  the possible numerical/physical uncertainties:
 
\begin{itemize}
\item {\it the turbulent kinetic energy} ranges from the case in which all velocity components larger than $L \geq 300 ~\rm kpc$ are filtered out (lower line), to the upper limit in which all the kinetic energy resolved in the cells is assumed to be turbulent. While our previous studies suggest that the typical outer scale of turbulent motions is of the order of $L \sim 100-300 ~\rm kpc$, in the presence of large-scale accretion this scale can increase. Hence the hatched region here is meant to bracket the plausible range of ICM turbulence. The turbulent energy support increases with radius, because the driving from infall motions in the outskirts is increasingly supersonic, due to the radial drop in the ICM temperature. 

\item {\it The magnetic energy} ranges from the primordial seed field model (trend increasing with radius) to the trend of the AGN seeding (trend decreasing with radius). In the AGN scenario the magnetic energy drops quickly with radius, as the overall activity by AGN is unable to significantly magnetise large volumes outside of clusters{\footnote{It shall be noted that also the underlying distribution of the thermal is slightly modified in the astrophysical seeding scenarios, due to the balance of cooling and feedback, which results into a denser cluster 
core.}}. 
Both these scenarios are probably underestimating the magnetic energy in the 
cluster centre, because the finite Reynolds number achieved in our run ($R_e \leq 500$, based on $R_e \sim 0.5~N^{4/3}$, where $N$ is the 1-dimensional size of the high-resolution domain of the simulations, in number of cells, e.g. \citealt{2006ApJ...638L..25K}) likely causes a delayed start of the small-scale dynamo amplification, compared to reality\citep[e.g.][]{2016ApJ...817..127B}. Limited to the single case of the Coma cluster, our primordial seeding run seems to be in better agreement with the observational results of Faraday Rotation \citep[e.g.][]{bo10,bo13}, which suggests a distribution of magnetic energy that scales with the gas thermal energy, and a significant magnetisation in the outskirts (even if limited to a single narrow sector of Coma, e.g. \citealt[][]{bo13});

\item {\it The cosmic-ray energy} goes from the upper limit obtained with out post-processing modelling of tracers \citep[][]{wi16} to zero in case no CR-protons are accelerated by shocks within the cluster.  The increasing trend with radius of the CR-energy follows from the sharp increase of the acceleration efficiency as a function of Mach number, which rapidly increase towards cluster outskirts \citep[e.g.][]{va10kp}. 

\end{itemize}

For comparison, we also show the upper and lower limits available for the different quantities by observations. For the turbulent motions, we show here the upper limits derived by \citet{2012MNRAS.421.1123C} from the analysis of X-ray fluctuations in the innermost region of the Coma cluster, which are at the level of $E_{\rm turb}/E_{\rm thermal} \sim 8\%$.  For the magnetic fields, we report the average ratio $E_{\rm mag}/E_{\rm thermal} \sim 1-2\%$ obtained through the modelling of Faraday Rotation in the Coma cluster by \citet{bo10} and \citet{bo13}. Finally, the ratio of CRs to the thermal gas energy has an upper limit based on the non-detection of hadronic $\gamma$-ray emission by \citet{fermi14}, of the order of $\sim 1-2\%$.  The simulated distribution of NT energy in this cluster seems compatible with 
observations, with the small exception of the magnetic energy, which we attribute to numerical dissipation.

The total budget of NT energy ($E_{\rm NT}=E_{\rm turb}+E_{\rm mag}+E_{\rm cr}$) becomes more relevant at large radii, as it ranges from $E_{\rm NT} \sim 3\% E_{\rm thermal}$ in the cluster core
to $E_{\rm NT} \sim 40\% E_{\rm thermal}$ at the cluster virial radius ($\sim 2.7 \rm ~Mpc$).
This radial trend is expected, as the virialisation of infall kinetic energy proceeds in an outside-in
fashion and is first mediated by outer accretion shocks, and later by the progressive dissipation of turbulent energy into gas heating (and possibly magnetic field amplification). 
In all cases the trend of NT energy is dominated by turbulence. In general, the lower bound given by our small-scale filtering should be considered probably the most realistic estimate of turbulence in the ICM in the central regions, while in cluster outskirts the presence of shock waves and infall motions makes this estimate more uncertain. 

For the other two NT components (cosmic rays and magnetic fields) the simulated trend suggests that their pressure contribution is small enough to consider them negligible for the dynamical evolution of the ICM. However, in the case of CR-energy this is the result of having assumed an acceleration efficiency of CR-protons which fulfils the comparison with FERMI limits, which are otherwise exceeded \citep[][]{scienzo16,wi16}. This shows that  despite the scarcity of direct observations of NT energy in the ICM, upper limits from observations are helpful not only to limit the NT energy budget of galaxy clusters nowadays, but also the underlying efficiency of processes responsible for the injection of NT energy in past epochs.

\subsection{The large-scale distribution of non-thermal energy.}

Using a second set of simulations, we generalise the previous results for the entire
cosmic volume. We must resort to simulations with a lower resolution, which cover
 a $50^3 ~\rm Mpc^3$ volume with $512^3$ cells for a {\it fixed} spatial resolution of $97 ~\rm kpc$, and $512^3$ dark matter particles. 
In this case, instead of using a tracer approach to model cosmic rays we use our direct simulation using a 2-fluid method we implemented in {\enzo} \citep[][]{scienzo,scienzo16}. 
However, this 2-fluid method does not work with the MHD solver in {\enzo}, and therefore 
we must combine independent resimulations of the same volume alternatively obtained including either CRs or magnetic fields. This approximated approach is valid as long as the two NT energies are small compared to the thermal energy and can therefore be viewed as two passive fields which do not affect the gas dynamics. Based on available constraints (see below) this assumption is probably reasonable. We will come back to this point in Sec.\ref{sec:discussion}.

The runs modelling cosmic rays have been produced with a  2-fluid model originally  developed in {\enzo} \citep[][]{scienzo,scienzo16}, which allow us to simulate at run-time the 
injection, the advection, the pressure feedback and the energy losses of CR-protons. The CRs are assumed to follow the ultra-relativistic equation of state ($\Gamma_{\rm cr}=4/3$) and are injected in the simulation by shock waves with an acceleration efficiency that scales with the Mach number \citep[][]{kr13}, rescaled by a factor $\sim 1/5$ to take into account the results of our recent work, where only quasi-parallel shocks are assumed to injected CR-protons \citep[][]{wi16}. We are also taking into account the effect of re-accelerated CRs in case of pre-existing CRs in the upstream of shocks \citep[][]{scienzo16}. 

The runs including magnetic fields use the MHD formulation using the Dedner cleaning (as in the previous Section) and were started either from the uniform primordial seed field ($B_0 = 10^{-9} \rm G$ comoving at $z=30$), or including the magnetic seeding from AGN ($1\%$ of the feedback thermal energy goes into magnetic energy) in simulations with radiative cooling. 

\begin{figure*}
\centering
\includegraphics{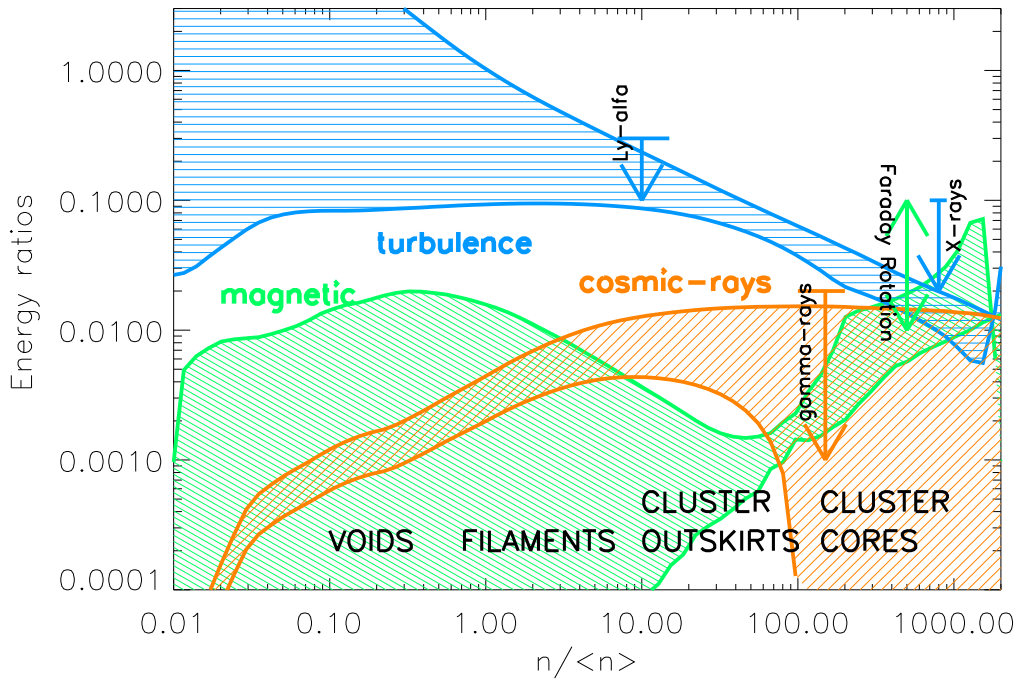}
\caption{Distribution of NT energies as a function of cosmic environment for a $50^3 ~\rm Mpc^3$ volume at $z=0$, normalised to the thermal energy within each density bin. The additional arrows show the available observational constraints for each energy field (see text for discussion).} 
\label{fig:NT}
\end{figure*}

From these simulations, starting from the same set of initial conditions, we computed  the distribution of thermal and non-thermal energies as a function of gas over density ($n/\langle n \rangle$, where $\langle n \rangle$ is the cosmic mean gas density) at $z=0$. For the NT energies we considered the turbulent energy, the CR-proton energy and the magnetic energy. 
Figure \ref{fig:NT} shows the distribution of these NT energy forms normalised to the thermal energy within the same density bins. 
Given the very large uncertainties in each of these quantities, we show through the various hatched regions in Fig. \ref{fig:NT} the 
range of uncertainties due to, both, numerical (e.g. resolution effects) and physical (e.g. acceleration efficiency of CRs, magnetic field seeds etc.) effects.\\

In detail, we show: 

\begin{itemize}
\item {\it the turbulent kinetic energy} is estimated as in the previous case via small-scale filtering (lower limit), or assuming that the entire post-shock kinetic energy is channeled into turbulence. This second option is particularly significant in the rarefied warm-hot intergalactic medium (WHIM) outside cluster, where our coarse resolution might underestimate vorticity 
injected by strong accretion shocks \citep[][]{ry08} . Again, the turbulent kinetic energy budget increases towards lower densities, becoming nearly supersonic at the scale of the linear structures of the Universe (e.g. filaments). At the over density probed through CIV absorption lines in the WHIM, the limits on shear motions at high-z are $\sim 10$ \% percent of the thermal energy \citep[][]{2001ApJ...554..823R}, which is at the level of our highest estimate of turbulence. For the cluster cores, the turbulent support is instead in line with present upper limits from X-ray  analysis by XMM-Newton observations  \citep[][]{sa11,2015A&A...575A..38P}, consistent with our estimate at higher resolution in the previous Section. 

\item  {\it The magnetic energy} is here measured by combining runs including only primordial seeding at high redshift, or with the additional seeding by AGN at run-time. Outside of clusters the uncertainties are very big because the fields are in the range $10^{-20} \rm G \leq B \leq 10^{-9}$ depending on the seeding model. Even more significantly than in the previous Section, our simulated magnetic fields are found to be smaller than what has been observed through Faraday Rotation \citep[e.g.][]{mu04,bo10,bo13}, due to the lack of resolution that prevents the formation of a small-scale dynamo. However, in the high density range of our distribution the magnetic energy can be significant, owing to the assumed fixed thermal/magnetic energy per event, which can have a higher impact on low mass systems compared to the previous analysis of the Coma-like cluster.

\item {\it The cosmic-ray energy} is limited by the requirement of staying within the $\gamma$-ray limits by FERMI \citep[][]{fermi14}.  While the uncertainties in the acceleration function produce a large uncertainty in the CR energy budget within clusters (owing to the fact that the acceleration efficiency of weak, $\mathcal{M} \leq 5$ shocks is very uncertain), the numerical uncertainties are smaller outside clusters, where shocks are predicted to be strong ($\mathcal{M} \gg 10$) and the acceleration efficiency is expected to be $\sim 10-30 \%$ based on the the modelling of strong supernova shocks \citep[e.g.][]{2012JCAP...07..038C}. 
\end{itemize}

Consistent with our previous analysis of the cluster, the support from NT energy increases moving away from structures and in less dense environments. Again, the kinetic turbulent support is found to be the dominant component, and also the one which is less affected by physical uncertainties. 
However, assessing the exact budget of turbulent pressure support in very rarefied environments is made difficult by the choice of an appropriate filtering scale, and also by the fact that the MHD  picture of this environment may eventually break down for very diluted plasmas.


\section{Discussion \& Conclusions}
\label{sec:discussion}

In this contribution we attempted to bracket the distribution of {\it all major components of NT energy} in galaxy clusters and in large-scale structures in general: turbulent kinetic energy, magnetic energy and cosmic-ray energy. 

Our simulations suggest that, everywhere in the cosmic volume, the NT energy from turbulent motions is dominant over other kinds of NT energies. The kinetic pressure support from turbulent motions  increases moving away from clusters and approaches the thermal budget at the scale of the linear structures of the Universe. 
This is  expected,  because  infall kinetic energy approaches the complete virialisation only at high over-densities, while in the more rarefied Universe turbulent motions, shocks and the additional role played by galactic outflows  can cause larger departures from virialisation and channel a larger energy into NT components. 
Consistent with observations,  the other two NT components (CR-energy and magnetic energy) 
are measured to reach the level of a few percent of the thermal energy budget in clusters.
While the first is highly uncertain due to the (unknown) acceleration efficiency of CRs by weak shocks, the distribution of magnetic energy is anchored to observations of Faraday Rotation in cluster cores. However, uncertainties in the efficiencies of dynamo amplification as well as on the origin of seed magnetic fields makes our predictions outside of clusters extremely uncertain.  
This makes any future observational attempts to detect these components appealing, because new detections (or even robust upper limits) have the potential of restricting the present range of uncertainties. In particular, detecting the continuum or polarised emission from shock-accelerated electrons in the WHIM will offer the chance of understanding, both, the origin of extragalactic magnetic fields and the  physics of structure formation shocks \citep[e.g.][]{va15radio,va15survey}.

Finally, a few important caveats to our analysis must be considered.
First, our simulated CRs are assumed to be frozen into the gas via coupling to the tangled magnetic fields lines. This is a sound enough approximation for the $\ge 10 ~\rm kpc$ scales considered here, but on long dynamical timescales the effect of CR diffusion might further smooth the distribution of CRs. 
Additionally, if CRs can stream at super-Alfvenic speed \citep[an interesting yet very debated scenario, e.g.][]{2011A&A...527A..99E} then the distribution derived in our approximation might be subject to further smoothing in the radial direction. 
Moreover, in our work we assumed that CRs and magnetic fields do not directly interact, while  more detailed work on CR-driven magnetic field instabilities suggest that this is not the case in small-scale features of the ICM \citep[e.g.][]{2013MNRAS.436..294B}. Finally, we also neglect the possible re-acceleration of CR-protons by turbulence \citep[e.g.][]{bj14}.

In the future,  constraining the level of
non-thermal energy in galaxy clusters (and especially in galaxy cluster outskirts) is crucial to perform high-precision cosmology. 
The expected NT budget in galaxy clusters is already constrained to be rather low.
X-ray and SZ analysis have shown that the NT pressure support is in general  $\leq  20 \%$ within $R_{\rm 200}$ in most clusters \citep[e.g.][]{2010ApJ...711.1033Z,2012MNRAS.425.2069M,2012MNRAS.425..162A,2013A&A...551A..22,2015ApJ...800...75F,2015MNRAS.450.2261M}.
Numerical simulations are important to predict the realistic 3-dimensional distribution of NT pressure and its dependence on the host cluster properties, such has total mass, formation epoch and dynamical state. 

To conclude, future cosmological simulations including all relevant NT components will have the potential to facilitate entering into an epoch of high-precision cosmology, and will also 
help future observations to become a physical probe of elusive processes in the very rarefied cosmic plasmas. 

\vspace{6pt} 
\acknowledgments{The computations described in this work were performed using the {\enzo} code (http://enzo-project.org), which is the product of a collaborative effort of scientists at many universities and national laboratories. We gratefully acknowledge the {\enzo} development group for providing extremely helpful and well-maintained on-line documentation and tutorials. The simulations were partially performed at the NIC of the Forschungszentrum J\"{u}lich, under allocations no. 9016 and 10755 (F.V. and D. W.) and no. 9059 (M. B.), and on Piz Daint (ETHZ-CSCS, Lugano) under allocation s585.  FV acknowledges financial support from the grant VA 876-3/1 by the Deutsche
153 Forschungsgemeinschaft (DFG). DW acknowledges support through grants SFB 676 and BR 2026/17 of the DFG.
F.V. and M.B. acknowledge partial financial support from the  FOR1254 Research Unit of the German Science Foundation (DFG). }

\authorcontributions{F. V. and D. W. performed the simulations used in this work; F. V. and M. B.  wrote this contribution; C. G. contributed to the development of 
the numerical tools necessary to analyse the data.}

\conflictofinterests{The authors declare no conflict of interest.}


\bibliographystyle{mdpi}
\bibliography{franco}

\end{document}